\documentclass[aps,pra,preprint,groupedaddress]{revtex4-2}

\usepackage{graphicx,caption,subcaption}
\newcommand{\eqref}[1]{(\ref{#1})}
\renewcommand{\varphi}{\wp} 
\begin{document}


\title{Non-existence of certain elliptic solutions of the Cubic-nonlinear Schr\"odinger Equation}


\author{Hans Werner Sch\"urmann}
\email[]{hwschuer@uos.de}
\affiliation{Department of Physics\\ University of Osnabr\"uck, Germany}

\author{Valery Serov}
\email[]{vserov@cc.oulu.fi,valserov@gmail.com}
\affiliation{Department of Mathematical Sciences\\ University of Oulu, Finland,\\
Moscow Centre of Fundamental and Applied Mathematics-\\ - Lomonosov Moscow State University, Russia}



\begin{abstract}
For a certain class of solutions of the cubic nonlinear Schr\"odinger equation we prove non-existence in the generic case. In the nongeneric case we present a two-parameter set of solutions, bounded or unbounded, depending on corresponding constraints.

\end{abstract}


\maketitle


\section{Introduction}

As is well known, elliptic (travelling-wave) solutions of the non-integrable complex Ginzburg-Landau equation
\begin{equation}
i\Psi_z(t, z)+\Psi_{tt}(t, z)+a\Psi(t, z)|\Psi(t, z)|^2-i\lambda\Psi(t, z)=0,\quad a\ne 0,\quad \lambda\in \mathbf{R},\quad \Psi\in \mathbf{C}
\end{equation}
do not exist \cite{H}, [2(a)]. If $\lambda=0$, (1) is integrable and coincides with the cubic nonlinear Schr\"odinger equation (CNLSE)
\begin{equation}
i\Psi_z(t, z)+\Psi_{tt}(t, z)+a\Psi(t, z)|\Psi(t, z)|^2=0,
\end{equation}
where, in the "fiber optics notation", $z$ denotes the distance along the fiber, and $t$ the (retarded) time. Besides general solutions derived by direct methods (e.g., IST, Darboux), particular solutions of the CNLSE, suitable for specific physical applications, are interesting and may suffice. In this context, methods have been proposed for seeking elliptic solutions of Eq.(1) [2(b)]. For the CNLSE particular elliptic  travelling-wave solutions exist in the form (see, e.g., \cite{S} and references therein)
\begin{equation}
\Psi(t, x)=f(z) e^{i(\phi(z)-\lambda t)},\quad z=x-ct.
\end{equation}
Thus, in correspondence to the non-existence of elliptic solutions of Eq.(1), the question is obvious whether elliptic solutions of Eq.(2), more general than those represented by (3), exist. In particular, if $f(z)e^{-i\lambda t}$ in (3) is replaced by $f(t, z)+id(z)$ (without assuming travelling-wave reduction $z=x-ct$), we are led to 
\begin{equation}
\Psi(t, z)=(f(t, z)+id(z))e^{i\phi(z)},\quad f,\phi, d\in \mathbf{R},
\end{equation} 
as a possible ansatz, suitable to obtain elliptic solutions of the CNLSE (2). 

If ansatz (4) is substituted in (2), we obtain, separating imaginary and real parts, the system 
$$
(a) \quad 
f_z(t, z)=d(z)(\phi_z(z)-a(d^2(z)+f^2(t, z))),
$$
\begin{equation}
(b)\quad 
f_{tt}(t, z)=d_z(z)+(\phi_z(z)-ad^2(z))f(t, z)-af^3(t, z),
\end{equation}
that must be assumed to hold necessarily, otherwise a study of its compatibility makes no sense.

To prepare the proof of inconsistency of system (5), in the next Section we derive solutions $d(z), \phi(z), f(t, z)$ by applying the Frobenius Theorem. Using these solutions, we present in Section III a numerical counterexample of Eq.(5a), rewritten as Eq.(14a) below, in the generic case. In Section IV we investigate the nongeneric case and present four solutions. Concluding Section V consists of a summary and of remarks, in particular concerning the physical relevance of the results presented.


\section{Tentative Solutions for the generic case $(d_z\ne 0, f_t\ne 0, f_z\ne 0)$}

In the following, we partly follow the line presented in [4(b)]: Equation (5b) admits a first integral 
\begin{equation}
(f_t(t, z))^2 = -\frac{a}{2} f^4(t, z) -(ad^2(z)-\phi_z(z)) f^2(t, z) + 2d_z(z) f(t, z) + 2b(z), 
\end{equation}
where $b(z)$ is a $t-$independent integration "constant". According to the Frobenius Theorem system $\{(5a),(6)\}$ has a (local) unique solution $f(t, z)$ if and only if the integrability condition $f_{zt}=f_{tz}$ is satisfied. Evaluation of the integrability condition yields a second degree polynomial in $f(t, z)$ whose vanishing coefficients are three ordinary differential equations
$$
(a)\quad \quad 4d(z)d_z(z)+\phi_{zz}(z)=0,
$$
$$
(b)\quad \quad b_z(z)+d(z)d_z(z)(\phi_z(z)-ad^2(z))=0,
$$
\begin{equation}
(c)\quad\quad d(z)(4ab(z)+(\phi_z(z)-ad^2(z))^2)+d_{zz}(z)=0.
\end{equation}
Equations (7a)-(7c) can be integrated successively, leading to ($c_1, c_2, c_3$ are integration constants)
$$
(a)\quad \quad \phi_{z}(z)=-2ad^2(z)+c_1,
$$
$$
(b)\quad \quad b(z)=\frac{1}{4}(2c_2-2c_1d^2(z)+3ad^4(z)),
$$
\begin{equation}
(c)\quad\quad (c_1^2+2ac_2)d^2(z)-4ac_1d^4(z)+4a^2d^6(z)+d^2_z(z)=c_3,
\end{equation}
where (8a) has been used to get (8b), and (8b) has been used for (8c). Setting $h(z)=d^2(z)$, Eq.(8c) can be rewritten as
\begin{equation}
(h_{z}(z))^2 = \alpha_1 h^4(z) + 4\beta_1 h^3(z) + 6\gamma_1 h^2(z) + 4\delta_1 h(z) + \epsilon_1 =: R_1(h),
\end{equation}
with
$$
\alpha_1 = - 16a^2,\quad \beta_1 = 4ac_1,\quad \gamma_1 = - \frac{1}{3}(2c_1^2 + 8ac_2),\quad \delta_1 = c_3,\quad \epsilon_1 = 0.
$$
Using (8a), (8b), Eq.(6) takes the form
\begin{equation}
(f_t(t, z))^2 = \alpha_2 f^4(t, z) + 4\beta_2 f^3(t, z) + 6\gamma_2 f^2(t, z) + 4\delta_2 f(t, z) + \epsilon_2 =: R_2(f, z),
\end{equation}
with
$$
\alpha_2 = - \frac{a}{2},\quad \beta_2 = 0,\quad \gamma_2 = \frac{1}{6}(c_1 - 3ah(z)),\quad \delta_2 = \frac{h_z(z)}{4\sqrt{h(z)}},\quad \epsilon_2 = 2c_2 + \frac{3}{2}ah^2(z) - c_1h(z).
$$
At this point, leaving the path described in [4(b)], we choose a well-known method to find solutions of Eqs.(9) and (10) by applying a formula due to Weierstrass \cite{W} : The nonlinear ODE
$$
(y_{x}(x))^2 = \alpha y^4(x) + 4\beta y^3(x) + 6\gamma y^2(x) + 4\delta y(x) + \epsilon =: R(y)
$$
is solved by (the prime denotes differentiation w.r.t. $y$)
$$
y(x)=y_0+
$$
\begin{equation}
+\frac{\frac{1}{2}R'(y_0)\left(\wp(x;g_{2},g_{3})-\frac{1}{24}R''(y_0)\right)\pm\wp'(x;g_{2},g_{3})\sqrt{R(y_0)}+\frac{1}{24}R(y_0)R'''(y_0)}{2\left(\wp(x;g_{2},g_{3})-\frac{1}{24}R''(y_0)\right)^2-\frac{1}{48}R(y_0)R''''(y_0)},
\end{equation}
where $y_0$ is an integration constant constant \cite{rem1}, and $g_2, g_3$ are the invariants of $R(y)$ \cite{KCh}. Applying (11) to (9) and (10), we get the elliptic solution
$$
h(z)=
$$
\begin{equation}
\frac{4\wp(z)(h_0\wp+\beta_1h_0^2+2\gamma_1h_0+\delta_1)+2\wp_z(z)\sqrt{R_1(h_0)}+
h_0^2(2\alpha_1\delta_1-2\beta_1\gamma_1)+h_0(4\beta_1\delta_1-5\gamma_1^2)-2\gamma_1\delta_1}{(2\wp(z)-\gamma_1-2\beta_1h_0-\alpha_1h_0^2)^2-\frac{\alpha_1}{2}R_1(h_0)},
\end{equation}
where $\wp(z)=\wp(z; g_{2h}, g_{3h})$ with invariants $g_{2h}, g_{3h}$ of $R_1(h)$
$$
g_{2h}=3\gamma_1^2-4\beta_1\gamma_1,
$$
$$
g_{3h}=-\gamma_1^3+2\beta_1\gamma_1\delta_1-\alpha_1\delta_1^2,
$$
and the doubly periodic solution $f(t, z)$ (elliptic in $t$, not elliptic in $z$)
$$
f(t, z; f_0(z))=
$$
\begin{equation}
\frac{-2\gamma_2\delta_2 - (5\gamma_2^2-\alpha_2\epsilon_2)f_0(z) + 2\alpha_2\delta_2 f_0^2(z) + 4\wp(t)(\delta_2+2\gamma_2f_0(z)+\wp(t)f_0(z)) + 
2\wp_t(t)\sqrt{R_2(f_0(z), z)}}{(2\wp(t)-\gamma_2-\alpha_2f_0^2(z))^2-\alpha_2R_2(f_0(z), z)}.
\end{equation}
Here $\wp(t)=\wp(t; g_{2t}, g_{3t})$ with invariants $g_{2t}, g_{3t}$ of $R_2(f, z)$
$$
g_{2t}=\frac{c_1^2}{12}-ac_2,\quad g_{3t}=\frac{ac_3}{8}-\frac{c_1(c_1^2+36ac_2^2)}{216},
$$
and $f_0(z)$ denoting a ($t-$independent) integration "constant".
Remarkably, coefficients in (10) are depending on $h(z)$, whereas $g_{2t}$ and $g_{3t}$ are $z-$independent.


With (8), system (5a), (6), can be rewritten as
$$
(a)\quad\quad f_z(t, z; f_0(z)) = \sqrt{h(z)}(c_1-a(3h(z)+f^2(t, z; f_0(z)))
$$
\begin{equation}
(b)\quad \quad (f_t(t, z))^2=R_2(f,z).
\end{equation}


\section{Inconsistency of System (14) in the generic case}





The problem of existence of elliptic solutions to CNLSE is reduced to the problem of solvability of system (14). We recall that Eq.(13) is necessary for the  existence of a solution of Eq.(2) by ansatz (4). With $h(z)$ given by (12), system (14) is solved by $f(t, z; f_0(z))$, if Eqs.(14a) \underline{and} (14b) are valid and compatible. Substituting $f$ into Eq.(14b), straightforward evaluation shows that Eq.(14b) is satisfied. If Eq.(14a) is \underline{assumed} to be valid also, then system (14) is compatible if $h(z)$ is given by Eq.(12). Hence (in the generic case) ansatz (4) is adequate if Eq.(14a) is satisfied by solution (13).
Substitution of $f(t, z; f_0(z))$ in (14a) is leading to a lengthy expression for $f_z(t, z; f_0(z))$ in terms of $\wp(z; g_{2h}, g_{3h}), \wp(t; g_{2t}, g_{3t}),$ and $\alpha_1, \beta_1,\gamma_1,\delta_1$, so that a validity check is highly complicated (even by using a computer algebra system). Instead, we present a counterexample by evaluating Eq.(14a) as 
$$
\Delta(t, z, f_0(z)):= f_z(t, z; f_0(z))-\sqrt{h(z)}(c_1-3ah(z)-af^2(t, z; f_0(z)))
$$
numerically with parameters
$a=-1, c_1=-2, c_2=0.4, c_3=0.03.$
Due to Eq.(13), $f$ depends on $z$ via coefficients $\alpha_2,\beta_2,\gamma_2,\delta_2,\epsilon_2,$ and on $f_0(z)=f(0, z)$. Studying the $t-$dependence of $\Delta$, we first choose $h_0=h(0)=0$, admissible with the parameters above. Second, we choose $z=1$, so that $f_0(1)$ must be selected admissibly as a zero of $R_2(f_0(1), 1)=0$ according to Eq.(10). Considering the dependence of $R_2(f_0(1), 1)$ on $f_0(1)$ (see Fig.1) and evaluating $f(t,1,f_{0i}(1)) (i=1,2,3,4)$, only $f_{02}$ and $f_{03}$ are associated to real and bounded $f$. Thus, by selecting $f_{03}=0.87$, Eq.(14a) as $\Delta(t,z,0.87)=0$ can be checked (by using MATHEMATICA, and confirmed by Maple). The result is shown in Fig.2:
Equation (14a) is violated for $z=1$, while Eq.(14b) is satisfied. Hence, ansatz (4) is not adequate to solve CNLSE (2) in the generic case $d_z(z)\ne 0$. -- The physical relevance of this result is considered in Section V.

\section{Nongeneric solutions}

Inconsistency of system (14) in the generic case indicates that condition $\{f_z\ne 0, f_t\ne 0, d_z\ne 0\}$ is too restrictive for the existence of a generic solution $\Psi(t, z)$. Thus it is appropriate to change the above condition, in order to check whether solutions exist subject to different conditions.

First, we consider the case $f_z\ne 0, f_t\ne 0$ and $d_z=0$. 
With $d(z)=const=:k$, system (8) reads.
$$
(a)\quad \quad \phi_z(z)=c_1,
$$
$$
(b)\quad \quad b(z)=c_2,
$$
\begin{equation}
(c)\quad \quad k(a^2k^4-2ac_1k^2+c_1^2+4ac_2)=0.
\end{equation}
Introducing $g(t, z)=f^2(t, z)$, system (5a), (6) can be rewritten as 
$$
(a)\quad \quad (g_z(t, z))^2=4k^2g(t, z)(c_1-ak^2-ag(t, z))^2,
$$
\begin{equation}
(b)\quad \quad (g_t(t, z))^2=2g(t, z)(4c_2+2(c_1-ak^2)g(t, z)-ag^2(t, z)).
\end{equation}
Due to (11), the solution of Eq.(16b) is given by
$$
g(t, z)=g_0(z)+
$$
$$
+3\frac{6\wp(t)\left(-3ag_0^2(z)+4(c_1-ak^2)g_0(z)+4c_2\right)+6\wp_t(t)\sqrt{-2ag_0^3(z)+4(c_1-ak^2)g_0^2(z)+8c_2g_0(z)}}
{(6\wp(t)+3ag_0(z)-2(c_1-ak^2))^2}+
$$
\begin{equation}
+3\frac{-3a^2g_0^3(z)+6a(c_1-ak^2)g_0^2(z)+(-12ac_2-8(c_1-ak^2)^2)g_0(z)+8c_2(ak^2-c_1)}
{(6\wp(t)+3ag_0(z)-2(c_1-ak^2))^2},
\end{equation}
with $\wp(t)=\wp(t;g_{2g}, g_{3g})$ and the invariants of (16b)
$$
g_{2g}=\frac{4}{3}(c_1^2+3ac_2),\quad g_{3g}=-\frac{4}{27}(2c_1^3+9ac_1c_2).
$$
Solution $g(t, z)$ depends on $z$ only via ($t-$independent) integration "constant" $g_0(z)$, that must (since $g_0(z)=g(0, z)$) satisfy system (16). Due to (16b), $g_0(z)$ is a constant $g_0$ so that 
$$
(a)\quad 4k^2g_0(c_1-ak^2-ag_0)^2=0,
$$
\begin{equation}
(b)\quad 2g_0(4c_2+2(c_1-ak^2)g_0-ag_0^2)=0
\end{equation}
must be solved subject to $k=0, k^2_{\pm}=\frac{c_1\pm2\sqrt{-ac_2}}{a}$ (according to (15c)) and $g_0\ge 0$. Reduction of (18) yields the admissible values $g_0$
$$
(a) \quad \text{if}\quad k=0:\quad g_0=0,\quad g_0=g_{0\pm}=\frac{c_1\pm\sqrt{c_1^2+4ac_2}}{a}>0,
$$
\begin{equation}
(b)\quad \text{if}\quad k^2=k^2_{-}:\quad g_0=0,\quad g_0=2\sqrt{-\frac{c_2}{a}},\quad ac_2<0
\end{equation}
(case $k^2=k^2_{+}$ must be excluded since it leads to $g_0=-2\sqrt{-\frac{c_2}{a}}<0$).

Considering system (16) with the possibilities due to (19), obviously it is satisfied if $k=0$. Using Eq.(17), the solutions are 

\begin{equation}
g(t)=g_1(t)=\frac{6c_2}{3\wp(t; g_{2g}, g_{g3})-c_1},
\end{equation}
\begin{equation}
g(t)=g_{\pm}(t)=4g_{0\pm}\frac{2c_1^2+9ac_2-3\wp(t; g_{2g}, g_{3g})(c_1+3\wp(t; g_{2g}, g_{3g}))}{(6\wp(t; g_{2g}, g_{3g})+c_1\pm3\sqrt{c_1^2+4ac_2})^2},\quad c_1^2+4ac_2\ge 0.
\end{equation}
Exploiting Eq.(17) with (19b), we obtain solutions identical with (20) and (21), if $k^2_{-}=0$. If $k^2_{-}\ne 0$, Eqs.(20), (21) do not satisfy Eq.(16a). In this case, system (16) is solved by 
\begin{equation}
g(t)=g_3(t)=-\frac{2c_2}{\sqrt{-ac_2}},\quad a>0,\quad c_2<0.
\end{equation}
We have obtained a two-parameter (due to Eq.(8c), we get $c_3=0$ for the foregoing solutions)
family of solutions $\Psi(t, z)$ according to ansatz (4), expressed in terms of elliptic function $\wp(t, g_{2g}, g_{3g})$ and $e^{ic_1z}$. Thus, $\Psi(t, z)$
is doubly periodic. The $t-$period $L_t$ is equal to the real period of $\wp(t, g_{2g}, g_{3g})$
\begin{equation}
L_t=2\omega(g_{2g}, g_{3g}).
\end{equation}
With (15a) the solutions $\Psi(t, z)$ are represented by  
\begin{equation}
\Psi(t, z)=\sqrt{g(t)}e^{i(c_1z+c_0)},
\end{equation}
with $g(t)$ given by (20)-(22), which must be real and non-negative (bounded or unbounded). To express this condition in terms of the parameters $a, c_1, c_2,$ it is appropriate to use a phase diagram approach \cite{ScSe} associated to solutions (20)-(21). According to Eq.(16b) (for (20)-(22), we must assume $k=0$) five and only five phase diagrams $\{g_t^2, g\}, g\ge 0,$ are possible and necessary for $g(t)$ to be real and non-negative, bounded or unbounded. Correspondingly, the parameters are constrained by


$$
(a):\quad \quad a<0,\quad c_1<0,\quad c_2>0,\quad c_1^2+4ac_2>0,
$$
$$
(b): \quad \quad a>0,\quad c_1>0,\quad c_2=0,
$$
$$
(c): \quad \quad a<0,\quad c_1<0,\quad c_1^2+4ac_2=0,
$$
$$
(d):\quad \quad a>0,\quad c_1\in \mathbf{R},\quad c_2>0,
$$
\begin{equation}
(e): \quad \quad a>0,\quad c_1>0,\quad c_2<0,\quad  c_1^2+4ac_2\ge0.
\end{equation}
If none of the constraints (25) are met, solutions (20) and (21) are not real and bounded (e.g., if $a<0, c_1<0, c_2<0,$ we get $g_1(t)<0, g_{+}(t)<0, g_{-}$ unbounded). If a certain constraint of (25) is valid, at least one solution according (20), (21) is bounded.
 
In cases $(a), (d)$ solutions $g_1$ and $g_{+}$ are bounded, $g_{-}$ is unbounded; in case $(e)$ solutions $g_{+}, g_{-}$ are bounded (different only by a shift in $t$), $g_1$ is unbounded; in case $(b)$, $g_{+}$ represents a bright solitary solution ($g, g_1= 0$ identically); in case $(c)$ $g_1$ represents a dark solitary solution ($g_{+}, g_{-}=\frac{c_1}{a}$). Solutions (20), (21), associated to $(b), (c)$, are degenerate elliptic, defined by $g^3_{2g}-27g^2_{3g}=0  (g_{3g}<0$ in both cases $(b),(c)$) \cite{Abr}. They are represented by
\begin{equation}
(b)\quad g_{+}(t)=\frac{2c_1}{a}sech^2(t\sqrt{c_1}),
\end{equation}
\begin{equation}
(c)\quad g_1(t)=\frac{c_1}{a}tanh^2\left(t\sqrt{-\frac{c_1}{2}}\right).
\end{equation}

To sum up, solution (24), with $g(t)$ given by (20), (21), (22), and subject to constraints (25), represents real and bounded as well as real, non-negative, and unbounded solutions of the CNLSE according to ansatz (4). -- Numerical evaluation is straightforward. Examples for real and bounded solutions are shown in Fig.3.

Second, we consider the case $f_t(t, z)=0$. -- The first integral $b(z)$ does not exist, so that we return to system (5):
$$
(a)\quad f_z(z)=d(z)(\phi_z(z)-a(d^2(z)+f^2(z))),
$$
\begin{equation}
(b)\quad d_z(z)+f(z)(\phi_z(z)-a(d^2(z)+f^2(z)))=0.
\end{equation}
Reduction of (28) yields 
\begin{equation}
d^2(z)+f^2(z)= const =c>0.
\end{equation}
If $d_z(z)=0 (f=const)$, $\phi(z)$ is given by 
\begin{equation}
\phi(z)=acz+c_0.
\end{equation}
If $d_z(z)=-\frac{f(z)f'(z)}{\sqrt{c-f^2(z)}}, |f(z)|< \sqrt{c},$ we obtain
\begin{equation}
\phi(z)=acz+\arcsin\left(\frac{f(z)}{\sqrt{c}}\right)+c_0.
\end{equation}
Compared with (24), (22), solution (30) represents no new solution. With respect to (31) we note that (29) and (31) satisfy (28), but, obviously, (29) and (31) are not sufficient to determine $f(z)$ and $\phi(z)$ in analogous manner like (17) and (15a). Nevertheless, (29), (31) are representing a solution according (4), whose physical relevance is unclear, however.

Third, if $f_t(t, z)\ne 0, f_z(t, z)=0$, no constraint for $d(z)$, the Frobenius Theorem cannot be applied, so that (again) system (5)
$$
(a)\quad d(z)(\phi_z(z)-a(d^2(z)+f^2(t)))=0,
$$
\begin{equation}
(b)\quad f_{tt}(t)=d_z(z)+f(t)(\phi_z(z)-a(d^2(z)+f^2(t)))
\end{equation}
must be reduced to yield 
$$
(a)\quad d(z)=0,\quad \phi_z(z)=\frac{f_{tt}(t)+af^3(t)}{f(t)},\quad f(t)\ne 0,
$$
\begin{equation}
(b)\quad \phi_z(z)=a(d^2(z)+f^2(t)),\quad d_z(z)=f_{tt}(t).
\end{equation}
From (33b) we get $f(t)=const, d_z(z)=0$, and thus, (again) a solution according to (22), (24). Considering (33a), we obtain that $\phi_z(z)=\frac{f_{tt}(t)+af^3(t)}{f(t)}=const=\lambda_1$ is necessary, leading to a linear function $\phi(z)$. Function $f(t)$ must satisfy ($\lambda_3$ denotes a further integration constant)
$$
(f_t(t))^2=-\frac{a}{2}f^4(t)+\lambda_1f^2(t)+\lambda_3,
$$
and hence (by setting $g(t):=f^2(t)$)
\begin{equation}
(g_t(t))^2=-2ag^3(t)+4\lambda_1g^2(t)+4\lambda_3g(t).
\end{equation}
Apart from the different designation of the integration constants, Eq.(34) is identical with Eq.(16a) if $k=0$. Solutions (20), (21) of (16b) are valid subject to $k=0$. Hence, solutions of Eq.(34) are given by (20), (21) (due to (15a), $\lambda_1$ is equal to $c_1$).

Finally we note that the degenerate solutions in the generic case, defined by  vanishing discriminants of $\wp(t; g_{2h},g_{3h})$ and $\wp(t; g_{2t},g_{3t})$, are nongeneric, but must be excluded as possible solutions due to the inconsistency in the generic case. Thus, apart from (31), nongeneric solutions are defined by $d_z(z)=0$ only.

\section{Conclusion}

\underline{(a)\quad Summary}\\
We have studied the adequacy of solution ansatz (4) (originally proposed in [4a]) for the CNLSE (2) as follows: \\ 
Assuming that ansatz (4) is adequate (as a solution of (2)) from (necessary valid) system (5) we have derived solutions $d(z),\phi(z), f(t, z; f_0(z))$ by exploiting the integrability condition $f_{zt}=f_{tz}$ of system $\{(5a), (6)\}$, and then used solutions $d, \phi, f$ to transform system (5) to system (14). If (14) can be satisfied by (13) (it is compatible if $h(z)$ is given by (12)), ansatz (4) is adequate. -- Due to the numerical counterexample presented in Section III, Eq.(14a) is not satisfied by Eq.(13), so that (4) is not adequate in the generic case. In the nongeneric case, using the results of Section II, we have derived solutions $\Psi(t, z)$ and corresponding constraints for non-negative bounded or unbounded solutions, respectively.



\underline{(b)\quad Remarks}\\ 
(i)\quad We emphasise that Eqs.(12), (13) do not represent a solution of the CNLSE (2). In the generic case, it is only necessary, and the reason to derive it, is its use for the counterexample in Section III. In Section IV Eqs.(12), (13) are applied to find the solutions in the nongeneric case.\\
(ii)\quad Equations (9), (10), (7a) and Eqs.(3.12), (3.14), (3.13) in [4(b)] are equivalent, respectively. Comparing the solution methods, it seems that using Weierstrass' formula (11) to solve (9) and (10) is more transparent and simple than the approach applied in Chapters 3.3, 3.4 in [4(b)].\\
(iii)\quad Even if we disregard the inconsistency of system (14), it should be emphasised that the corresponding test of consistency of the Riccati equation (3.4) [4(b)] and Eq.(3.14) was not even considered in [4(b)].\\
(iv)\quad The physical relevance of the result of Section III is 
due to the general importance of the CNLSE in physics as well as to the fact that solutions in 
the seminal article [4(a)] have found numerous citations and applications, in particular in optics  and hydrodynamics \cite{CMKTA}-\cite{ChPW}. Thus it is appropriate that some results in the literature should be reconsidered.
To be specific,  we exemplarily refer to \cite{VSN}, where the doubly periodic background solutions presented in \cite{CMKTA} (on the basis of [4a]; see [48] in \cite{CMKTA})
are compared with experimental data. The authors of \cite{VSN} claim "good agreement  between theory and experiment".  Since we are not competent enough to assess the details of the experimental setup, we note that only the zero-order and first-order Fourier coefficients of $Q(t, z)$ (see Eq.(16) in \cite{VSN}) are compared with the data (see Figs.3 and 4 in \cite{CMKTA}) during only two periods of $t$ ("two cycles of evolution" \cite{CMKTA}). Furthermore, the discrepancies between theory and data (see hatched regions in Figs.3 and 4) beyond the two cycles are ascribed (ad hoc) to imperfect loss compensation. --  To sum, we suggest to improve the  "good" agreement between theory and experiment by using the correct ansatz (4) --  solutions of the CNLSE.\\
(v)\quad A hydrodynamical application of solutions in [4] is presented in  \cite{ChPW}, where rogue waves and modulation instability of the wave background are modelled by the focusing CNLSE using Eqs.(2) and (3) in \cite{ChPW}, that are derived from (60) (identical with (3.23) in [4b])
and (65) (identical with (3.26) in [4(b)] and from (69), (75), respectively. Apart from different definitions of $z$ and $t$ the solutions (2) and (3) in \cite{ChPW} are not consistent with Eqs.(20) and (21) above, since $g_1(t), g_{\pm}(t)$ are $z-$independent while the amplitude functions in (2) and (3) are dependent on both variables. To put it simply in a different way: $Q(x, t)$ according to Eq.(65) is a solution of (51), first equation (with $\delta^2$ according to (60) and $\Theta'=-2(\delta^2+b)$), but it is an unsolved problem whether it is a solution of the second equation of system (51). As long as this problem is not solved, it seems  doubtful that article \cite{ChPW} "opens up a number of new directions in the study of rogue waves modelled by the focusing  NLS equation". \\
(vi) \quad In a recent article \cite{C} new elliptic ansatz (4) -- solutions of the CNLSE have been presented. The solutions $Q(x, t)$ of Eq.(5) in \cite{C} (with $h(t)$ according to Eq.(7) in \cite{C}), derived by using Eq.(11) above, does not satisfy Eq.(13) in \cite{C}.



\begin{figure}[h!]
\includegraphics[width=14cm]{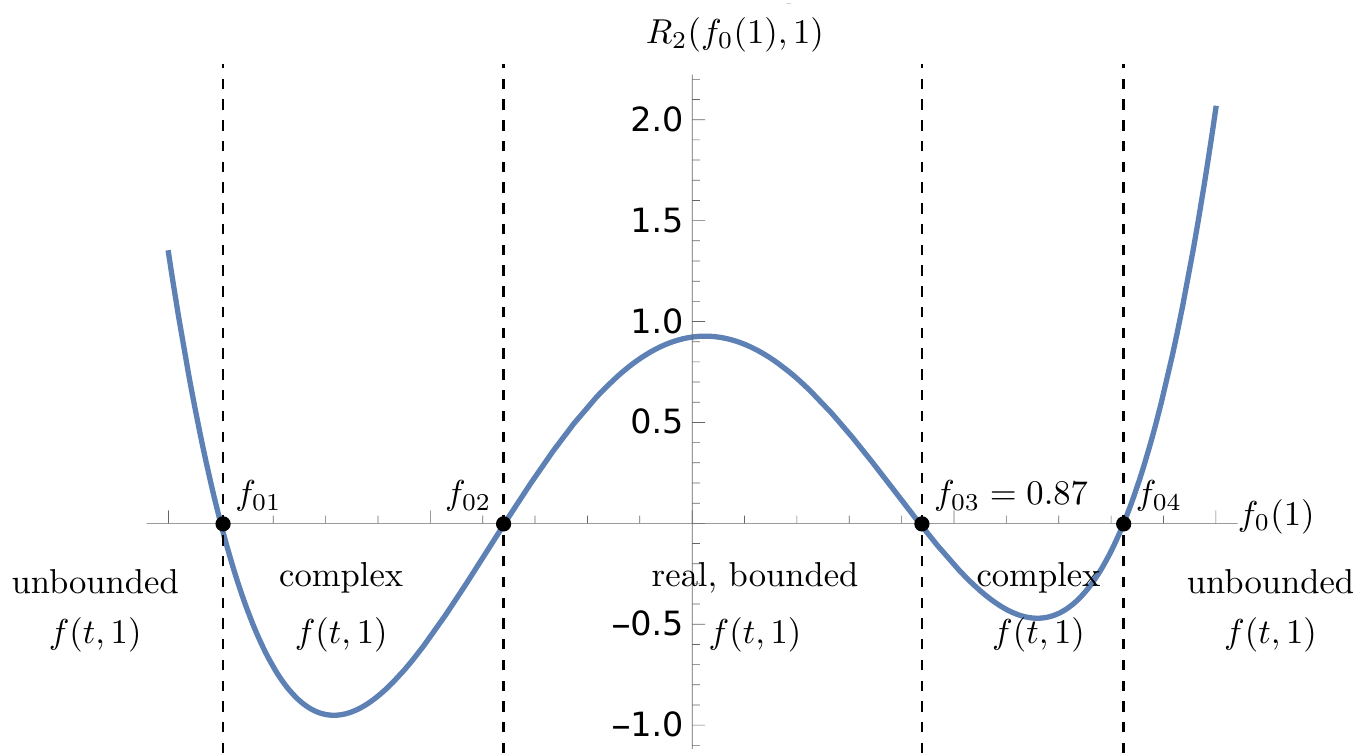}
\caption{Phase diagram Eq.(10) for $z=1$, parameters $a=-1, c_1=-2, c_2=0.4, c_3=0.131$ (see the text).}
\end{figure}

\begin{figure}[h!]
\includegraphics[width=14cm]{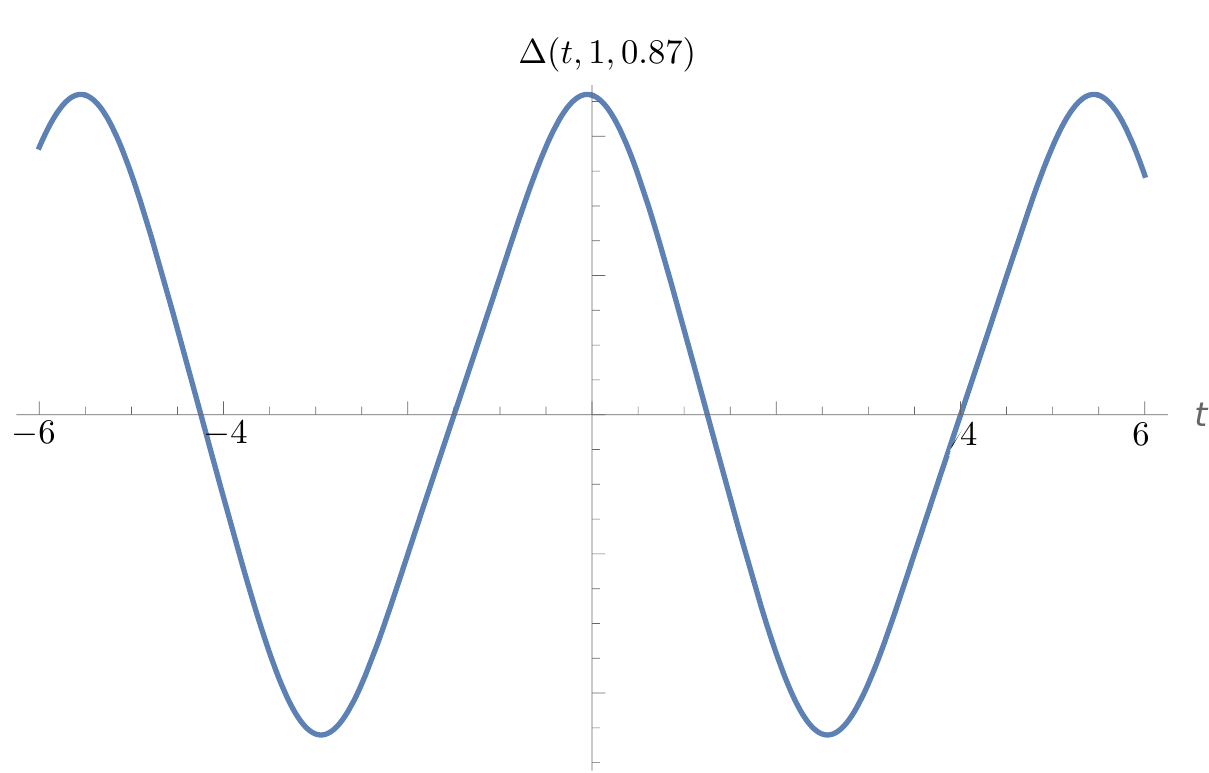}
\caption{Test of Eq.(14a) $(\Delta(t, 1, 0.87)\ne 0)$ for $z=1$, parameters as in Fig.1}
\end{figure}

\begin{figure}[h!]

\begin{subfigure}[b]{0.47\textwidth}
         \centering
         \includegraphics[width=\textwidth]{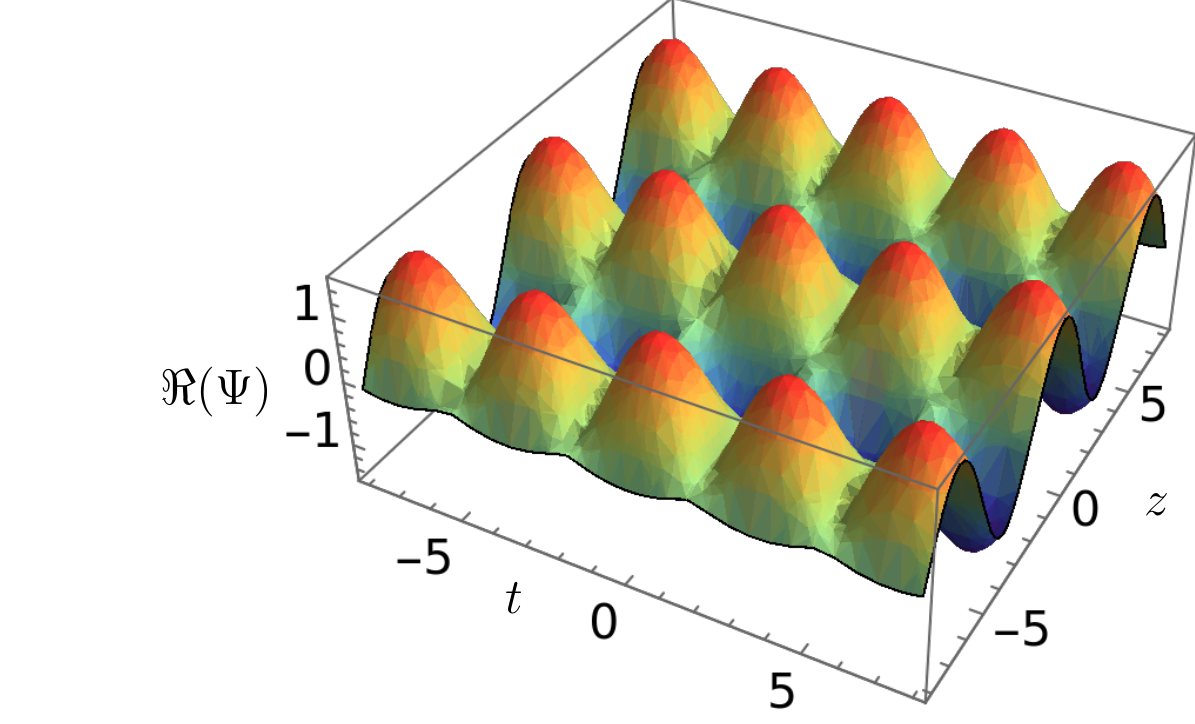}
         \caption{}
         \label{fig3a}
\end{subfigure}
\begin{subfigure}[b]{0.47\textwidth}
         \centering
         \includegraphics[width=\textwidth]{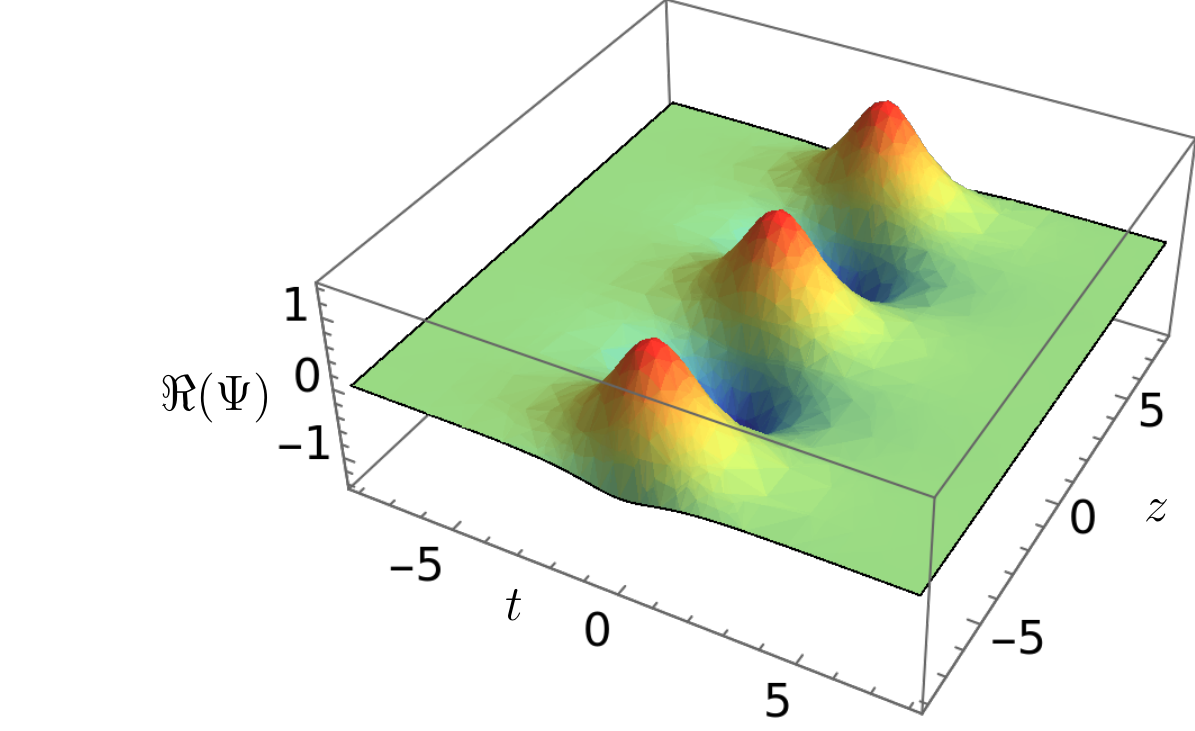}
         \caption{}
         \label{fig3b}
\end{subfigure}
\begin{subfigure}[b]{0.47\textwidth}
         \centering
         \includegraphics[width=\textwidth]{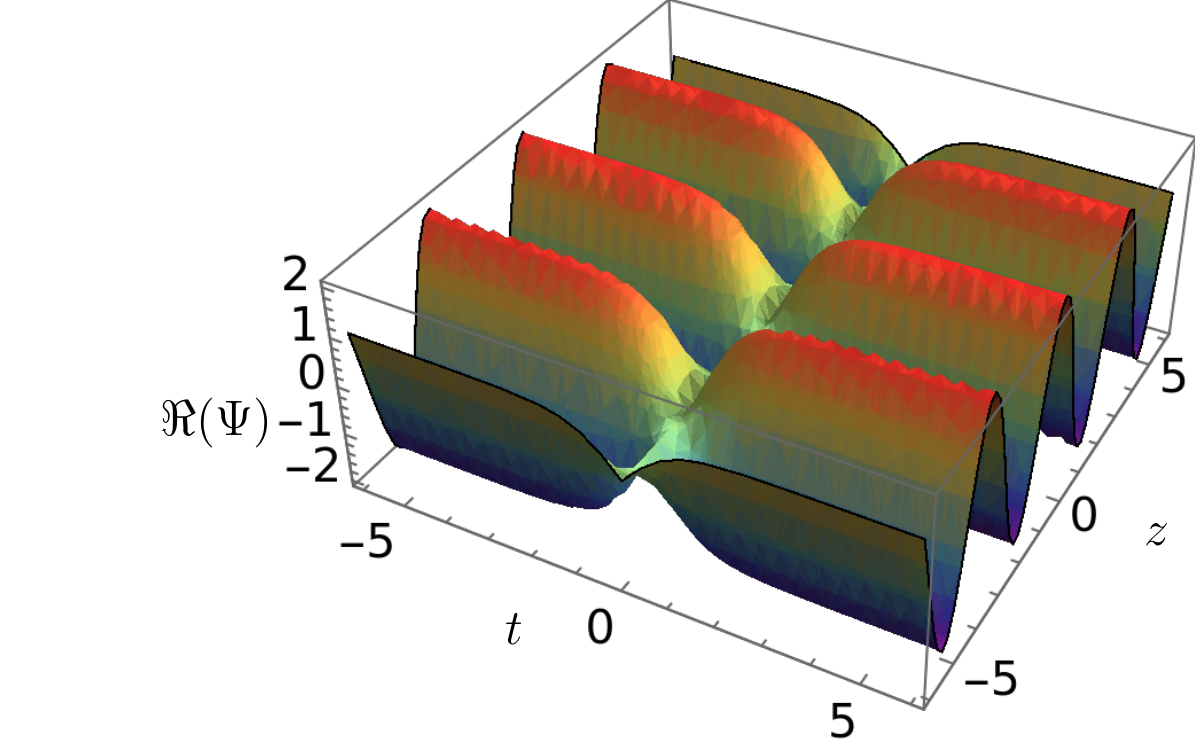}
         \caption{}
         \label{fig3c}
\end{subfigure}

\caption{$\Re (\Psi(t, z))$ according to (24), parameters according to constraints $(a), (b), (c)$ in (25): (a): $g_{+}(t)$ with $a=-\frac{1}{8}, c_1=-1, c_2=1$; (b):  $g_{+}(t)$ with $a=1, c_1=1, c_2=0$; (c): $g_1(t)$ with $a=0.46, c_1=-1.92, c_2=2.$}
\end{figure}






\clearpage

\section*{References}



\begin{thebibliography}{99}
\bibitem{rem1} Since (11) implies $y(0)=y_0, R(y_0)=0$ holds; but (11) is valid for any constant $y_0$, not necessarily for a zero of $R(y)=0$.
\end{thebibliography}
\end{document}